\documentclass[letterpaper]{article} 
\usepackage{aaai2026}  
\usepackage{times}  
\usepackage{helvet}  
\usepackage{courier}  
\usepackage[hyphens]{url}  
\usepackage{graphicx} 
\usepackage{natbib}  
\usepackage{caption} 
\usepackage{algorithm}
\usepackage{algorithmic}
\usepackage{amsmath}
\usepackage{amssymb}
\usepackage{booktabs}
\usepackage{tabularx}
\usepackage{pifont}
\usepackage{multirow}
\usepackage{makecell}
\usepackage{array}
\usepackage{tabularx}
\usepackage{enumitem}
\newcommand{\cmark}{\ding{51}}
\newcommand{\xmark}{\ding{55}}

\usepackage{newfloat}
\usepackage{listings}
\DeclareCaptionStyle{ruled}{labelfont=normalfont,labelsep=colon,strut=off} 
\floatstyle{ruled}
\newfloat{listing}{tb}{lst}{}
\floatname{listing}{Listing}
\nocopyright

\title{CTR-Driven Ad Text Generation via Online Feedback Preference Optimization}

\author{
    Yanda Chen\textsuperscript{\rm 1}\equalcontrib,
    Zihui Ren\textsuperscript{\rm 2}\equalcontrib,
    Qixiang Gao\textsuperscript{\rm 2},
    Jiale Chen\textsuperscript{\rm 2},
    Si Chen\textsuperscript{\rm 2},
    Xubin Li\textsuperscript{\rm 2},
    Tiezheng Ge\textsuperscript{\rm 2}\thanks{Corresponding author.},
    Bo Zheng\textsuperscript{\rm 2}
}

\affiliations{
    \textsuperscript{\rm 1}Harbin Institute of Technology, Shenzhen\\
    \textsuperscript{\rm 2}Taobao \& Tmall Group of Alibaba\\
    cydaaa30@gmail.com\\
    \{renzihui.rzh, gaoqixiang.gxq, chenjiale.cjl, chensi.cs, lixubin.lxb, tiezheng.gtz, bozheng\}@alibaba-inc.com
}

\usepackage{bibentry}

\begin{document}

\maketitle

\begin{abstract}
Advertising text plays a critical role in determining click-through rates (CTR) in online advertising.
Large Language Models (LLMs) offer significant efficiency advantages over manual ad text creation.
However, LLM-generated ad texts do not guarantee higher CTR performance compared to human-crafted texts, revealing a gap between generation quality and online performance of ad texts.
In this work, we propose a novel ad text generation method which optimizes for CTR through preference optimization from online feedback.
Our approach adopts an innovative two-stage framework:
(1) diverse ad text sampling via one-shot in-context learning, using retrieval-augmented generation (RAG) to provide exemplars with chain-of-thought (CoT) reasoning;
(2) CTR-driven preference optimization from online feedback, which weighs preference pairs according to their CTR gains and confidence levels.
Through our method, the resulting model enables end-to-end generation of high-CTR ad texts.
Extensive experiments have demonstrated the effectiveness of our method in both offline and online metrics.
Notably, we have applied our method on a large-scale online shopping platform and achieved significant CTR improvements, showcasing its strong applicability and effectiveness in advertising systems.
\end{abstract}

\section{Introduction}
Advertising text is crucial in online advertising, appearing in almost every ad.
Effective ad texts not only feature well-crafted expression and accurate conveyance of item information~\cite{murakami2022aspect}, but also achieve strong performance in terms of click-through rate (CTR).
Large Language Models (LLMs) enable automated ad text generation (ATG)~\cite{mita2023striking} from basic item information and simple instructions, offering significantly higher efficiency than manual creation.
Some works~\cite{golobokov2022deepgen,chai2022fast,jin2023towards,kamigaito2024generating} focus on offline metrics like diversity or fluency to enhance the quality of generated ad texts.
However, these methods do not guarantee higher CTR performance compared to human-crafted texts, revealing a gap between offline metrics optimization and online deployment effectiveness.

\begin{figure}[h!]
    \centering
    \includegraphics[width=0.45\textwidth]{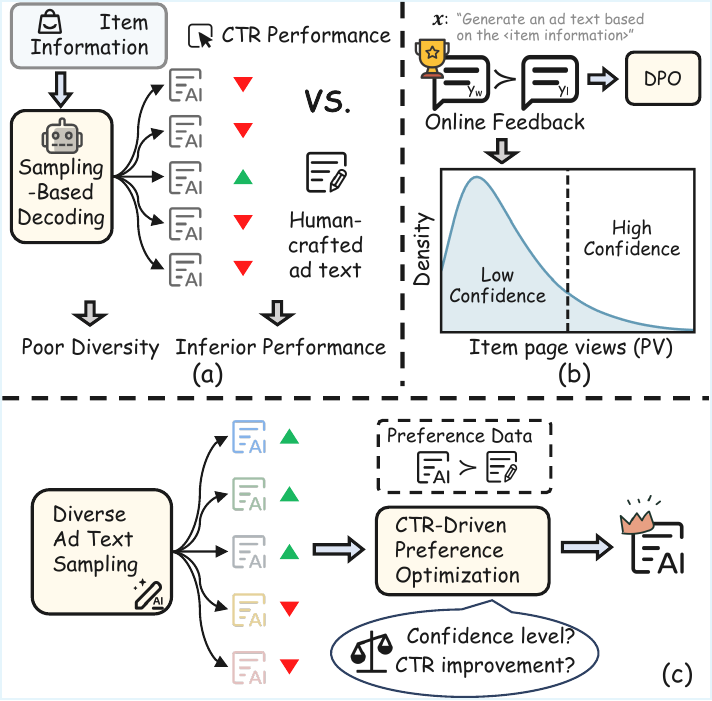}
    \caption{\textbf{Challenges in high-CTR ad text generation.} (a) Poor diversity and inferior CTR performance of sampling-based decoding strategy. (b) Noisy online feedback due to uneven page view ($PV$) distribution across items. (c) We propose an innovative two-stage framework: (1) \textbf{diverse ad text sampling} and (2) \textbf{CTR-driven preference optimization} tailored for these challenges.}
    \label{fig:teaser}
\end{figure}

Recently, studies~\cite{kanungo2022cobart, wei2022creater, murakami2025adparaphrase} have focused on directly optimizing for online metrics, particularly CTR. 
CTR has long been regarded as a key metric for evaluating online advertising effectiveness, as clicks serve as the prerequisite for subsequent conversion to other revenue-related metrics.
These approaches typically generate multiple ad texts via sampling-based decoding from LLMs for online testing.
The generation models are subsequently optimized using online feedback through algorithms such as contrastive learning, semi-supervised learning, and reinforcement learning.

However, these approaches often suffer from the low quality of sampling-based decoding and the noise in online feedback, making it difficult to achieve a significant CTR advantage over human-crafted ad texts.
We conduct some analytical experiments and reveal the challenges to realize the CTR improvement.
Figure~1(a) shows that generating multiple ad texts via sampling-based decoding from LLMs results in poor diversity and inferior CTR performance.
Figure~1(b) illustrates data confidence issues arising from the uneven distribution of item page views ($PV$) in online feedback.
Short-term online testing makes it difficult to accumulate sufficient high-PV data, causing substantial noise in online feedback.
These challenges constrain further improvements in CTR optimization of ATG tasks.

In this work, we propose CTOP, a novel \underline{\textbf{C}}TR-driven ad \underline{\textbf{T}}ext generation method via \underline{\textbf{O}}nline feedback \underline{\textbf{P}}reference optimization.
CTOP employs an innovative two-stage framework consisting of diverse ad text sampling and CTR-driven preference optimization for high-CTR ad text generation, as illustrated in Figure~1(c).
In the first stage, we design a \textbf{diverse ad text sampling} strategy to address the low quality of sampling-based decoding.
Specifically, we generate candidate ad texts via a style transfer task, imitating the style of high-quality exemplars. 
For each item, the strategy uses retrieval-augmented generation (RAG) to retrieve multiple chain-of-thought (CoT)-augmented ad text exemplars, each serving as a one-shot in-context example.
In the second stage, we conduct \textbf{CTR-driven preference optimization} tailored for noisy online feedback.
We select winning candidates that outperform their corresponding human-crafted ad texts in online A/B/n testing and construct preference pairs by treating the former as preferred and the latter as dispreferred.
Each pair is assigned a training weight calculated from gain and confidence coefficients based on its CTR improvement and reliability.

Extensive experimental results on offline and online metrics demonstrate the effectiveness of CTOP.
Our method enables efficient candidate generation of diverse high-quality ad texts and better aligns with online user feedback during preference optimization.
Notably, we have successfully applied our method to the item title generation task on a large-scale online shopping platform across various advertising display scenarios.
Since late 2024, our approach has achieved a relative increase of 1.11\% in click-through rate (CTR) and 1.02\% in revenue per mille (RPM) compared to the human-crafted baseline, showcasing its strong applicability and effectiveness in advertising systems.
The main contributions of our work can be summarized as follows:
\begin{itemize}
    \item We propose a novel ad text generation method which optimizes for CTR via preference optimization from online feedback, closing the gap between generation quality and online performance of ad texts.
    \item We design an innovative two-stage framework, consisting of (1) diverse ad text sampling via one-shot in-context learning, using RAG to provide high-quality exemplars with CoTs; and (2) CTR-driven preference optimization method that dynamically weighs training pairs by gain and confidence coefficients based on online feedback.
    \item We demonstrate the practicality and effectiveness of our method through comprehensive offline and online experiments, and show its successful deployment on a large-scale online shopping platform, achieving significant improvements in CTR and RPM (revenue per mille).
\end{itemize}

\section{Related Works}
\subsection{Advertising Text Generation}
Advertising Text Generation (ATG)~\cite{murakami2023natural,mita2023striking} aims to automate the creation of high-quality ad texts.
Early template-based methods~\cite{bartz2008natural,thomaidou2013automated} lack diversity and scalability, while extractive approaches~\cite{thomaidou2013grammads} ensure content consistency but struggle to generate linguistically varied or user-engaging outputs.

Later works increasingly treat ATG as a sequence-to-sequence task, using pre-trained LLMs to generate fluent, tailored ad texts.
These methods explicitly optimize for key dimensions: content diversity~\cite{golobokov2022deepgen,chai2022fast,jin2023towards}, fluency~\cite{zhang2021chase,wei2022creater,li2022culg}, faithfulness~\cite{youngmann2020automated,shao2021controllable,golobokov2022deepgen}, and relevance to user intent~\cite{kamigaito2021empirical,duan2021query}.
In this work, we directly optimize for the performance metric—click-through rate (CTR)—as the primary objective of ATG.

\subsection{CTR-Oriented Ad Text Generation}
Recently, online performance—particularly click-through rate (CTR)—has emerged as a central objective in ATG research~\cite{wang2019quality,mishra2020learning,kamigaito2021empirical,kanungo2022cobart,wei2022creater,murakami2025adparaphrase}, as it directly reflects user engagement and advertising effectiveness. 
Typically, these studies first generate diverse ad texts via sampling-based decoding for online A/B testing.
Then, different algorithms are leveraged to learn from online feedback, such as reinforcement learning~\cite{hughes2019generating,kamigaito2021empirical,kanungo2022cobart}, contrastive learning~\cite{wei2022creater}, semi-supervised learning~\cite{shao2021controllable}, and training-free in-context learning~\cite{murakami2025adparaphrase}.

However, these methods suffer from the low-quality sampling-based decoding and the noisy online feedback, leading to insufficient alignment with user preference.
In our work, we design diverse ad text sampling and CTR-driven preference optimization tailored for these issues.

\subsection{Reinforcement Learning in Advertising Creativity}
With the success of Reinforcement Learning from Human Feedback (RLHF)~\cite{christiano2017deep,ziegler2019fine,ouyang2022training,ji2023ai,wang2024comprehensive}, RL has been widely used in ad creativity tasks like text and image generation.
Early works~\cite{hughes2019generating,kamigaito2021empirical,kanungo2022cobart} adopt self-critical RL~\cite{rennie2017self} to optimize predicted CTR (pCTR) or proxy metrics like ad quality scores.
Recent works in ad image generation~\cite{yang2024new,chen2025ctr} train CTR reward models on user preferences and use them to guide PPO~\cite{schulman2017proximal} or label preference data for DPO~\cite{rafailov2023direct}.
In our work, we propose CTR-driven preference optimization based on the DPO algorithm to optimize ATG for CTR improvement.

\section{Task Formulation and Challenges}
\subsection{CTR-driven Ad Text Generation}
We formulate the Ad Text Generation (ATG) task with click-through rate (CTR)—the core online metric—as the optimization objective. 
Given the item information $x$, the goal is to learn a parameterized generator $G_\theta$ that maximizes the expected CTR of generated ad texts $y$:

\begin{equation}\label{eq1}
\theta^* = \arg\max_\theta \mathbb{E}_{x \sim \mathcal{X},\, y \sim G_\theta(\cdot|x)} \left[ \mathrm{ctr}(y) \right],
\end{equation}
where $\mathrm{ctr}(y) = \frac{\mathrm{click}(y)}{\mathrm{pv}(y)}$ is computed from online clicks and ad page views ($PV$).

However, evaluating whether a generated ad text achieves ``high CTR'' is challenging without a meaningful baseline. 
In practice, each item has a human-crafted ad text $y_\text{hu}$ typically written by merchants or the advertising platforms based on domain expertise or empirical knowledge.
These human-crafted ad texts generally achieve a certain level of quality and CTR, which serve as a strong and interpretable baseline.
Therefore, any generated ad text $y$ that outperforms $y_\text{hu}$ in online testing can be considered achieving high CTR.
To this end, we reformulate the objective as maximizing the CTR improvement over human-crafted ad texts:

\begin{equation}\label{eq3}
\theta^* = \arg\max_\theta \mathbb{E}_{x \sim \mathcal{X},\, y \sim G_\theta(\cdot|x)} \left[ \mathrm{ctr}(y) - \mathrm{ctr}(y_\text{hu}) \right].
\end{equation}

This relative formulation is more tractable to optimize and better business-aligned than the absolute form in Equation~1.
It calls for the effective generation of candidate ad texts with high-CTR potential to outperform human-crafted baselines during online testing.

\begin{figure}[h!]
    \centering
    \includegraphics[width=0.45\textwidth]{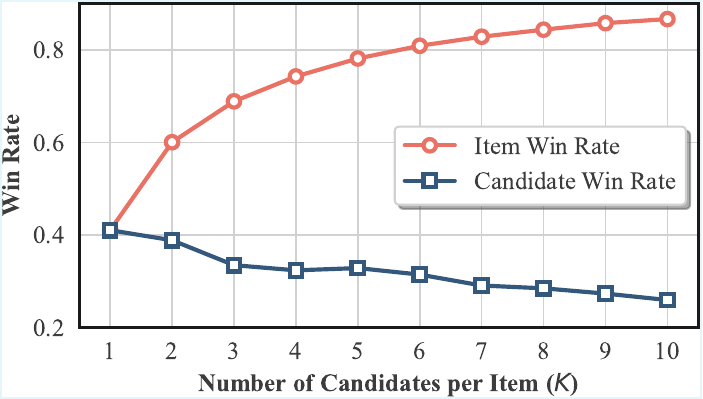}
    \caption{Item and candidate win rates over human-crafted ad texts with different numbers of candidates per item ($K$).}
    \label{fig:llm-sampled win rate}
\end{figure}

\begin{table}[h!]
\small
\centering
\begin{tabular}{@{}ccccc@{}}
\toprule
\multirow{2}{*}{\textbf{\#Cand. ($K$)}} 
& \multicolumn{2}{c}{\textbf{Self-BLEU~$\downarrow$}}
& \multicolumn{2}{c}{\textbf{Distinct-$n$~$\uparrow$}} \\
\cmidrule(lr){2-3} \cmidrule(lr){4-5}
& $n=1$ & $n=2$ & $n=2$ & $n=3$ \\
\midrule
5  & 0.540 & 0.376 & 0.422 & 0.712 \\
10 & 0.680 & 0.497 & 0.342 & 0.650 \\
\bottomrule
\end{tabular}
\caption{Diversity metrics of ad texts generated by sampling-based decoding ($K=5$ and $K=10$ candidates per item).}
\label{tab:diversity_llm_sampling}
\end{table}

\subsection{Challenge I: Low Quality of Sampling-based Decoding}
We select the item title generation task on a large online shopping platform to explore generating high-CTR candidate ad texts.
Given the flexibility of linguistic expression, introducing diversity into candidate ad texts is a natural strategy to generate more high-CTR candidates.

Sampling-based decoding methods, such as top-$k$ and top-$p$ sampling~\cite{holtzman2020curious}, are widely adopted to generate diverse outputs.
We use the pre-trained Qwen-2.5-7B~\cite{qwen2.5} to generate $K$ candidate ad titles per item via top-$k$ sampling ($temperature=1.5$, $top\_k=100$).
Both the candidate and human-crafted titles are deployed on the platform for online A/B/n testing, where titles of an item are shown to users with equal probability.
We conduct a two-week A/B/n test on 1 million items and perform statistical analysis, with results shown in Figure~2.

We define the \emph{item win rate} as the proportion of items for which at least one candidate title achieves a higher CTR than the human-crafted.
As shown in Figure~2, the item win rate increases with $K$, indicating that introducing diversity contributes to generating more high-CTR ad texts.

However, the \emph{candidate win rate}—the fraction of individual candidates that outperform the human-crafted title—decreases as $K$ increases.
This inferior CTR performance indicates that most generated titles are of low quality. 
We further assess candidate diversity, computing both Self-BLEU and Distinct-$n$ scores for the generated titles. 
As shown in Table~1, both metrics suggest that increasing $K$ leads to greater redundancy and diminishing diversity.

Sampling-based decoding restricts generation to high-probability tokens but still inherits model biases and risks output degradation, especially at high temperatures.
The poor diversity and inferior CTR performance reveal the low quality of sampling-based decoding.

\begin{figure}[b!]         
    \centering
    \includegraphics[width=0.45\textwidth]{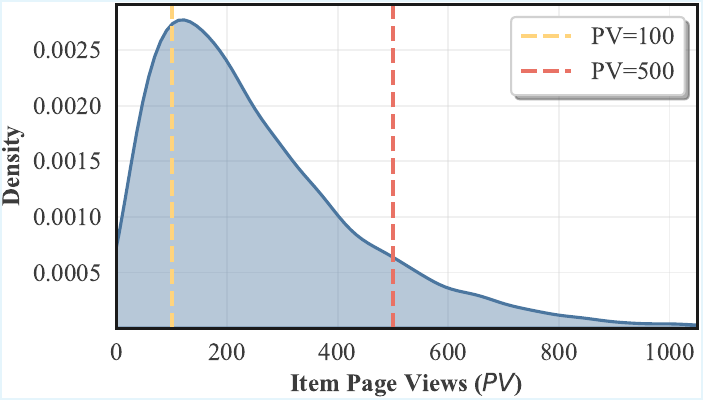}
    \caption{Distribution of item page views ($PV$) after two weeks of online testing.}
    \label{fig:pv distribution}
\end{figure}

\begin{figure*}[htbp]     
    \centering
    \includegraphics[width=\textwidth]{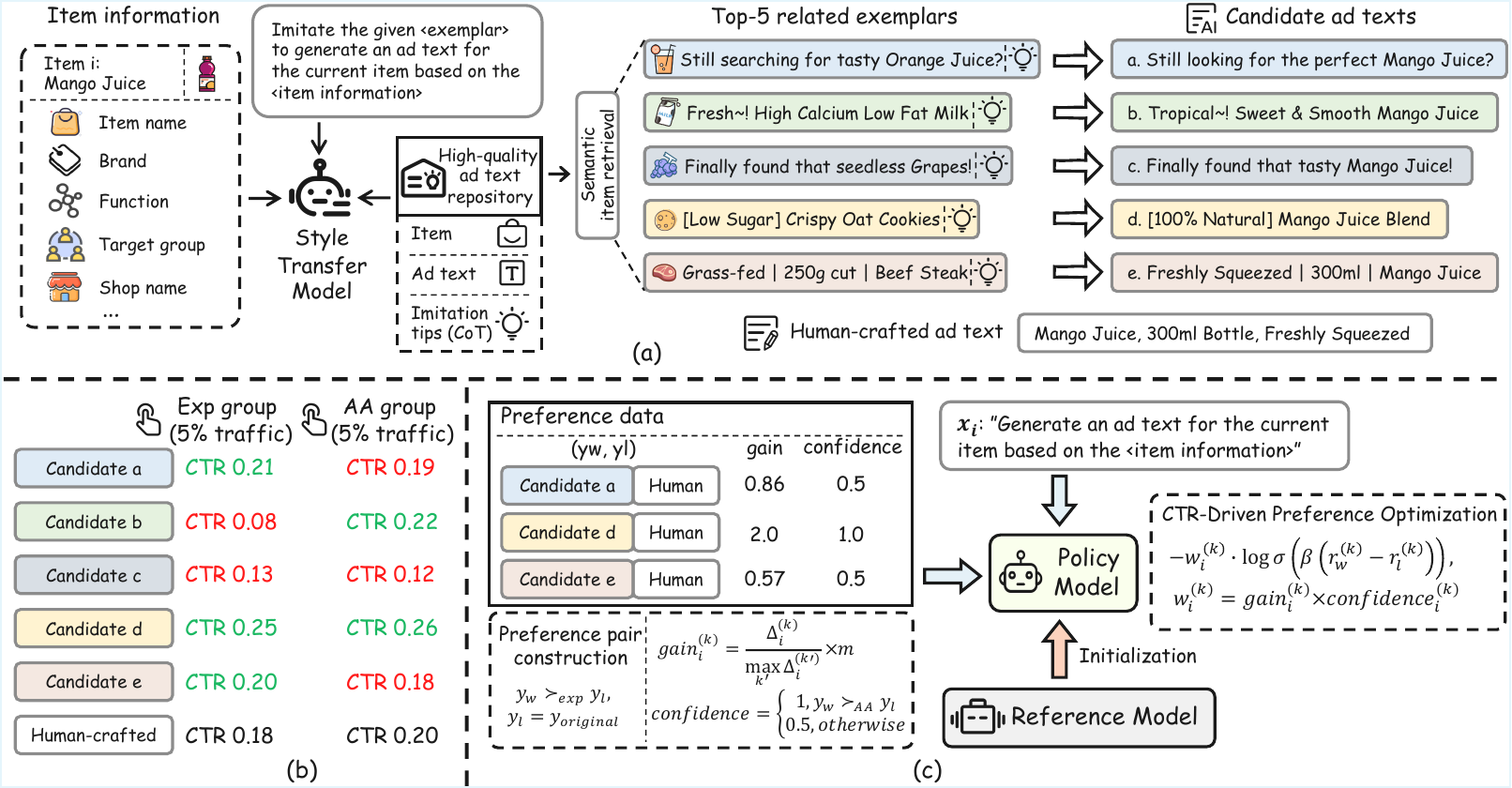}
    \caption{Overall framework of our CTR-driven ad text generation method, CTOP. (a) \textbf{diverse ad text sampling} via one-shot in-context learning, using RAG to provide exemplars with CoT reasoning. (b) \textbf{Online A/B/n testing} and a parallel control \textbf{AA group} with identical settings for confidence measurement. (c) \textbf{CTR-driven preference optimization} weighing preference pairs according to their CTR \textbf{gain} and \textbf{confidence} levels.}
    \label{fig:arch}
\end{figure*}

\begin{table}[h]
\centering
\begin{tabular}{@{}cccc@{}}
\toprule
\textbf{$PV$ Level} & Low  & Medium  & High  \\
\midrule
\textbf{Pair Accuracy (\%)} & 52.3 & 54.9 & 57.6 \\
\bottomrule
\end{tabular}
\caption{Pair accuracies of DPO models trained with preference data of different item page view ($PV$) levels.}
\label{tab:pv_dpo}
\end{table}

\subsection{Challenge II: Optimization under Noisy Feedback}
Optimizing ad text generation for high CTR aligns naturally with reinforcement learning (RL), where the goal is to maximize user engagement in real-world environments. 
In our setting, online A/B/n testing yields an off-policy dataset comprising both LLM-generated and human-crafted texts.
Moreover, CTR feedback is delayed, noisy, and requires significant online accumulation, making it challenging to train an explicit reward model for CTR prediction or preference estimation.
These factors motivate us to adopt off-policy RL methods with implicit reward modeling~\cite{rafailov2023direct,pal2402smaug,wu2024beta,azar2024general,kim2024sdpo,ethayarajh2024kto,liu2024lipo}.

Here we adopt DPO~\cite{rafailov2023direct} due to its simplicity and efficiency. DPO operates on preference pairs $(x, y_w, y_l)$, where $y_w$ and $y_l$ are the preferred and dispreferred responses with input context $x$. 
Given a policy model $\pi_\theta$ and a reference model $\pi_\mathrm{ref}$, the DPO objective is:
\begin{equation}\label{eq4}
\small
\mathcal{L}_{\text{DPO}} = - \log \sigma \left( 
    \beta \left[ 
        \log \frac{ \pi_\theta(y_w \mid x) }{ \pi_{\text{ref}}(y_w \mid x) }
        - 
        \log \frac{ \pi_\theta(y_l \mid x) }{ \pi_{\text{ref}}(y_l \mid x) }
    \right]
\right)
\end{equation}
where $\beta$ denotes the preference strength, and $\sigma$ is the sigmoid function.

We construct preference pairs with each winning candidate as $y_w$ and its human-crafted ad text as $y_l$ from online feedback. 
A Qwen-2.5-7B model is fine-tuned on diverse candidate ad texts as $\pi_\mathrm{ref}$. 
With a dataset of 50K pairs from online A/B/n testing, we train a DPO-optimized model. 
However, A/B testing against human-crafted ad texts yields only a 42.1\% win rate.

We attribute this to the low confidence of preference pairs, especially for items with limited page views ($PV$). 
To investigate, we conduct an offline study by splitting the preference data into three PV-level subsets: low ($PV \leq 100$), medium ($100 < PV \leq 500$), and high ($PV > 500$). 
A separate high-PV test is constructed to evaluate pairwise accuracy:

\begin{equation}
\text{Acc} = \frac{1}{N} \sum_{i=1}^N \mathbb{I}\left[\pi_\theta(y_w^{(i)} \mid x^{(i)}) > \pi_\theta(y_l^{(i)} \mid x^{(i)})\right]
\end{equation}

As shown in Table 2, preference data with higher $PV$ yields better DPO performance due to its higher confidence. 
However, Figure~3 presents the uneven item $PV$ distribution after two weeks of online testing, with only a small fraction reaching high $PV$.
Training data from noisy online feedback limits the effectiveness of preference optimization.

\section{Method}
In this section, we introduce CTOP, a novel \underline{\textbf{C}}TR-driven ad \underline{\textbf{T}}ext generation method via \underline{\textbf{O}}nline feedback \underline{\textbf{P}}reference optimization.
To ensure meaningful improvements in online performance, we formulate the task objective as maximizing the CTR improvement over human-crafted ad texts, defined in Equation~2.
As illustrated in Figure~4, our method employs an innovative two-stage framework of diverse ad text sampling and CTR-driven preference optimization tailored for the two challenges.
To address the low-quality issue of sampling-based decoding, we design a \textbf{diverse ad text sampling} strategy in the first stage.
To better learn from noisy online feedback, we perform \textbf{CTR-driven preference optimization} in the second stage.
The following subsections elaborate on each stage of our proposed method.

\subsection{Diverse Ad Text Sampling}
As depicted in \textbf{Challenge I}, candidate ad text generation via sampling-based decoding causes poor diversity and inferior CTR performance.
To effectively generate candidates of both high quality and true diversity, we design a diverse ad text sampling strategy, shown in Figure~4(a).
The core of our strategy is to formulate candidate generation as a style transfer task, where the model imitates high-quality exemplars via one-shot in-context learning.

To support this, we first construct a repository of high-quality ad text exemplars.
These exemplars are either crafted by domain experts or selected from successful online cases with proven strong CTR performance.
Each exemplar consists of the item information, the ad text, and a chain-of-thought (CoT) with imitation tips to guide faithful style transfer.
We pre-generate offline CoTs for all exemplars in the repository for efficiency.

Then, we train a style transfer model tailored for the ad text style transfer task.
The model generates candidate ad texts via one-shot in-context learning.
For each target item, our strategy uses retrieval-augmented generation (RAG) to retrieve the top-$K$ semantically relevant exemplars from the repository based on item information.
Each exemplar serves as a one-shot in-context example, guiding the style transfer model to generate one stylistically imitated candidate ad text, yielding $K$ diverse outputs.

Owing to the inherent stylistic diversity among high-quality exemplars in the repository, our strategy is able to maintain both quality and diversity even as the number of candidates per item $K$ increases.
Moreover, this diversity is controllable: our strategy can generate new styles of ad texts simply by collecting additional exemplars without extra training.
Compared to methods that demand manual prompt engineering for each new style, our strategy offers a more scalable and efficient solution with minimal overhead.

\subsection{CTR-Driven Preference Optimization}
Following our diverse ad text sampling strategy, we deploy the generated candidates and the human-crafted ad texts for short-term online A/B/n testing, as shown in Figure~4(b).
A candidate that achieves a higher CTR than its corresponding human-crafted ad text is defined as a \emph{winning candidate}.
We construct preference pairs with each winning candidate as $y_w$ and its human-crafted ad text as $y_l$.
However, our analysis in \textbf{Challenge II} reveals that noisy online feedback negatively impacts the quality of preference optimization.
To address this issue, we propose a CTR-driven preference optimization approach illustrated in Figure~4(c), where each pair is assigned a dynamic weight calculated from \textbf{gain} and \textbf{confidence} coefficients.

The gain coefficient measures the improvement of the winning candidate’s CTR over its human-crafted ad text.
We first compute the absolute CTR difference for each preference pair within the same item.
Let $\Delta_i^{(k)} = \mathrm{ctr}\big(y_w^{(k)}\big) - \mathrm{ctr}\big(y_l^{(k)}\big)$ denote the CTR difference within the $k$-th pair of item $i$.
To account for varying baseline CTRs across items and ensure numerical stability, we normalize $\Delta_i^{(k)}$ by the maximum CTR difference within the same item:
\begin{equation}\label{eq6}
\mathrm{gain}_i^{(k)} = \frac{\Delta_i^{(k)}}{\max_{k'} \Delta_i^{(k')}} \times m
\end{equation}
where $m$ is a scaling factor that amplifies small differences for better distinguishability in preference optimization.

The confidence coefficient quantifies the reliability of each pair, based on the relative order of the pair between the experimental group and the parallel control group with identical settings.
As shown in Figure~4(b), the parallel control group mirrors the experimental group setup exactly, including ad content and traffic allocation.
In online advertising, such a parallel control group is commonly referred to as the \textbf{AA group}, serving to evaluate the stability of ad performance.
For each preference pair $(y_w, y_l)$, we compare its relative order in both the experimental group and the AA group for confidence calculation:

\begin{equation}\label{eq7}
\mathrm{confidence} = 
\begin{cases}
1, &  y_w \succ_{\text{exp}} y_l \cap y_w \succ_{\text{AA}} y_l, \\
0.5, & \text{otherwise}.
\end{cases}
\end{equation}

This binary confidence score ensures that stable and reproducible preferences contribute fully to model training, while mitigating the impact of unreliable pairs.

For each preference pair $k$ of item $i$, we assign a dynamic weight $w_i^{(k)} = \mathrm{gain}_i^{(k)} \cdot \mathrm{confidence}_i^{(k)}$.
Building upon DPO~\cite{rafailov2023direct} algorithm, our final CTR-driven preference optimization objective is formulated as follows:

\begin{equation}\label{eq:ctrdpo}
\mathcal{L}_{\text{CTRPO}} = -\sum_{i=1}^N \sum_{k=1}^{M_i} w_i^{(k)} \cdot \mathcal{L}_{\text{DPO}}
\end{equation}
where $N$ is the total number of items, $M_i$ is the number of preference pairs constructed for item $i$, and $\mathcal{L}_{\text{DPO}}$ denotes the DPO loss as formulated in Equation~3.

Our approach enables prioritizing preference signals of both substantial CTR improvement and high reliability, leading to more robust and effective ad text generation.

\section{Experiments}
In this section, we primarily evaluate our method through online experiments, supplemented by offline studies.
We first compare our proposed method with other baselines via short-term online A/B testing with human-crafted ad texts.
Next, we conduct experiments to validate the impact of the two stages in the framework, respectively.
Finally, we present the long-term online business gains by deploying our method at scale on a large-scale online shopping platform.

\subsection{Experiment Setup}
\subsubsection{Implementation Details}
We conduct online experiments on a large-scale e-shopping platform, with the task of generating item titles across display advertising scenarios.
For diverse ad text sampling, we first build a high-quality item title repository with offline CoTs.
A Qwen-2.5-7B model is trained on the ad text style transfer task as the style transfer model.
Then we retrieve the top-5 relevant exemplars per item and use each as a one-shot in-context example for style transfer.
We select 200K items across different categories for diverse candidate generation, resulting in a total of 1 million candidate ad titles.
These candidates, together with the human-crafted item titles, are deployed for a two-week online A/B/n testing using 5\% of platform traffic as the experimental group.
A parallel AA group is simultaneously deployed with identical content and traffic allocation.

For CTR-driven preference optimization, we first construct a preference dataset of 500K pairs by pairing each winning candidate and its corresponding human-crafted title from the online A/B/n testing.
Each pair is assigned a dynamic weight based on its gain coefficient (Equation~5, with scaling factor $m=2.5$) and confidence coefficient (Equation~6).
We use a Qwen-2.5-7B model fine-tuned on diverse candidate titles as the reference model $\pi_{\text{ref}}$, which also initializes the policy model $\pi_\theta$.
We optimize the objective in Equation~7 for one training epoch with a learning rate of $2 \times 10^{-5}$.  
All training processes are conducted on machines equipped with 8 NVIDIA A100 GPUs.

Due to business sensitivity concerns, we are unable to publicly release the data used in the experiments. See Appendix for more implementation details.

\subsubsection{Baselines}
For fair comparison, we evaluate our method against the following baselines with the same online data by our diverse ad text sampling strategy.
All baseline models are implemented with Qwen-2.5-7B.
\begin{itemize}
    \item \textbf{Prompt-only}: Directly generating titles from item information using simple prompts without extra model tuning.
    \item \textbf{Top-1 sampling}: Our diverse ad text sampling with sampled number $K=1$ (i.e., the most related exemplar).
    \item \textbf{Top-1 SFT}: Supervised fine-tuning on the highest-CTR winning candidate of each item from the A/B/n testing. 
    \item \textbf{DPO}: The standard Direct Preference Optimization (DPO) algorithm~\cite{rafailov2023direct}, applied to preference pairs $(\text{winning candidate}, \text{human-crafted})$.
\end{itemize}

\subsubsection{Evaluation Metrics}
We select 1 million items for evaluation, ensuring that none of them overlap with those in the training set.
For each method, a single ad title is generated for each item then deployed alongside its corresponding human-crafted title for a two-week A/B testing on the online shopping platform, using 5\% of total traffic.
The following metrics are calculated from the online feedback:
\begin{itemize}
    \item \textbf{Win Rate}: The proportion of items where the generated title achieves a higher CTR than the human-crafted title.
    \item \textbf{Overall Relative CTR Improvement (CTR~$\uparrow$)}: The relative improvement in average CTR of the generated titles over the human-crafted titles, aggregated across all items.
\end{itemize}

\begin{table}[htbp]
\centering
\begin{tabular}{@{}lcc@{}}  
\toprule
\textbf{Method} & \textbf{Win Rate (\%)} & \textbf{CTR~$\uparrow$ (\%)} \\
\midrule
\multicolumn{3}{l}{\textit{w/o preference alignment}} \\
Prompt-only & 41.0 & -9.83 \\
Top-1 sampling & 48.7 & -0.6 \\
\midrule
\multicolumn{3}{l}{\textit{w/ preference alignment}} \\
Top-1 SFT & 52.3 & +0.92 \\
DPO & 56.0 & +3.48 \\
\textbf{CTOP (Ours)} & \textbf{60.2} & \textbf{+4.76} \\
\bottomrule
\end{tabular}
\caption{Comparison of online A/B testing results across different methods. Our method outperforms all baselines.}
\label{tab:online_results}
\end{table}

\begin{figure}[b!]
    \centering
    \includegraphics[width=0.45\textwidth]{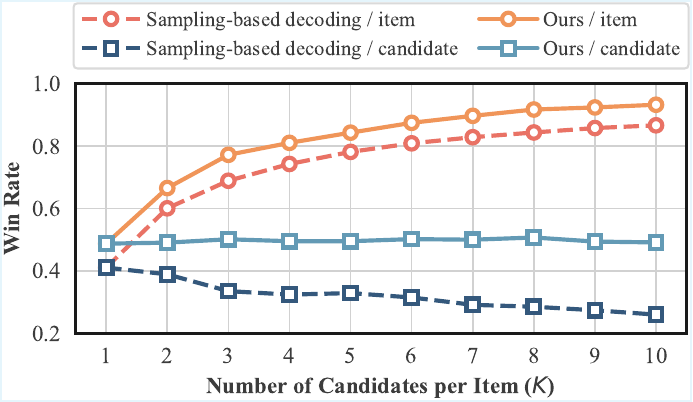}
    \caption{Win rate (over human-crafted ad texts) comparison between sampling-based decoding and our diverse ad text sampling with different numbers of candidates ($K$).}
    \label{fig:win rate comparison}
\end{figure}

\begin{figure*}[htbp]     
    \centering
    \includegraphics[width=\textwidth]{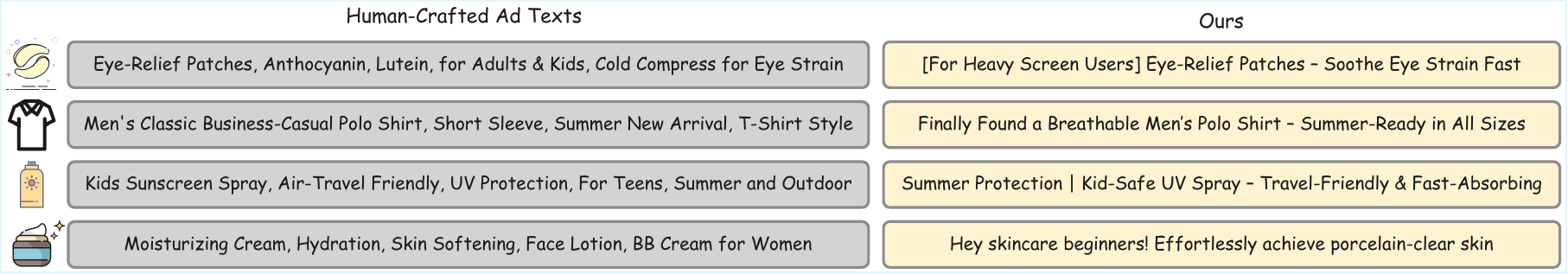}
    \caption{Comparison cases between human-crafted item titles and our generated titles on the online shopping platform.}
    \label{fig:case study}
\end{figure*}

\subsection{Main Results}
Table 3 shows the online A/B testing results. Our method achieves the best win rate (60.2\%) and relative CTR improvement (+4.76\%), outperforming all baselines and demonstrating a significant positive impact on overall CTR.

For baselines without preference alignment (\textbf{Prompt-only} and \textbf{Top-1 sampling}), both underperform the human-crafted ad texts. 
\textbf{Prompt-only} yields a low win rate (41.0\%) and negative CTR improvement (-9.83\%), revealing the gap between generation quality and online performance of ad texts.
\textbf{Top-1 sampling} improves win rate through only the diverse ad text sampling strategy, but still results in a slight CTR reduction (-0.6\%). 
This suggests that style transfer does not ensure suitability for the target item, underscoring the need for preference optimization from online feedback.

In contrast, preference alignment methods leverage online feedback for model optimization, all achieving higher win rates ($>$50\%) over the human-crafted baseline.
Among these methods, \textbf{DPO} outperforms \textbf{Top-1 SFT}, demonstrating the advantage of reinforcement learning to align with user engagement.
Our method \textbf{CTOP} extends DPO by introducing dynamic weights of preference pairs based on CTR gain and confidence, achieving superior performance.
Through prioritizing high-confidence, high-gain pairs, the model learns from the most meaningful feedback, leading to more robust and effective ad text generation.

\begin{table}[htbp]
\small
\centering
\begin{tabular}{@{}cc|cc|cc@{}}
\toprule
\multirow{2}{*}{\makecell{\textbf{\#Cand.}\\($K$)}} & \multirow{2}{*}{\textbf{Method}}
& \multicolumn{2}{c|}{\textbf{Self-BLEU~$\downarrow$}}
& \multicolumn{2}{c}{\textbf{Distinct-$n$~$\uparrow$}} \\
\cmidrule(lr){3-4} \cmidrule(lr){5-6}
& & $n=1$ & $n=2$ & $n=2$ & $n=3$ \\
\midrule
\multirow{2}{*}{5}  & \makecell{Sampling-based\\decoding} & 0.540 & 0.376 & 0.422 & 0.712 \\
                    & \textbf{Ours}         & \textbf{0.454} & \textbf{0.277} & \textbf{0.648} & \textbf{0.930} \\
\midrule
\multirow{2}{*}{10} & \makecell{Sampling-based\\decoding} & 0.680 & 0.497 & 0.342 & 0.650 \\
                    & \textbf{Ours}         & \textbf{0.599} & \textbf{0.389} & \textbf{0.555} & \textbf{0.890} \\
\bottomrule
\end{tabular}
\caption{Comparison of diversity metrics between sampled-based decoding and our diverse ad text sampling ($K=5$ and $K=10$ candidates per item).}
\label{tab:diversity comparison}
\end{table}

\subsection{Analysis on Diverse Ad Text Sampling}
The comparison between \textbf{Prompt-only} and \textbf{Top-1 sampling} in the main results (Table 3) has demonstrated the effectiveness of our diverse ad text sampling strategy.
To provide a further analysis, we comprehensively compare our method with sampling-based decoding ($temperature=1.5, top\_k=100$) across different candidate numbers $K$.

As shown in Figure~5, our method consistently outperforms sampling-based decoding on both item and candidate win rates across all $K$.
Furthermore, while the candidate win rate of sampling-based decoding drops with larger $K$ (from 41.0\% to 25.9\%), our method remains stable at around 50\%, indicating strong quality of the individual candidate.

We also assess the diversity metrics in Table 4. 
Our method produces lower self-BLEU and higher Distinct-$n$ values than sampling-based decoding, demonstrating the ability to generate diverse and high-quality candidates.

\begin{table}[h]
\centering
\small
\begin{tabular}{@{}cc|cc@{}}
\toprule
\textbf{Confidence} & \textbf{Gain} & \textbf{Win Rate (\%)} & \textbf{CTR~$\uparrow$ (\%)} \\
\midrule
\xmark & \xmark & 56.0 & +3.48 \\
\cmark & \xmark & 58.5 & +3.79 \\
\cmark & \cmark & \textbf{60.2} & \textbf{+4.76} \\
\bottomrule
\end{tabular}
\caption{Ablation results of gain/confidence coefficients in CTR-driven preference optimization.}
\label{tab:ablation_prefopt}
\end{table}

\subsection{Ablation on CTR-driven Preference Optimization}
We conduct an ablation to evaluate the impact of gain and confidence coefficients in our CTR-driven preference optimization. 
As shown in Table 5, removing both coefficients reduces the method to standard DPO, which yields a 56.0\% win rate and +3.48\% relative CTR improvement. 
Adding only the confidence coefficient improves the win rate to 58.5\% and CTR to +3.79\%, indicating better stability by prioritizing reliable preference pairs. 
Our full method, which incorporates both gain and confidence, further boosts the win rate to 60.2\% and relative CTR improvement to +4.76\%. 
These results demonstrate that jointly considering both factors produces the best overall performance: gain emphasizes the impact of preference pairs on CTR improvement, while confidence mitigates noise from unreliable data.

\subsection{Online Results}
We deployed our proposed CTOP for item title generation on a large-scale e-shopping platform across various ad scenarios.
In initial A/B/n testing, the candidate ad texts showed a -0.46\% CTR decrease versus the human-crafted baseline, indicating the necessity of preference alignment.
After CTR-driven optimization, the generated titles achieved a relative improvement of +1.11\% in CTR and +1.02\% in RPM (revenue per mille) over the human-crafted baseline under full platform traffic, since late 2024.
The significant performance improvement demonstrates the strong applicability and effectiveness of CTOP in real-world advertising systems.

\section{Case Study}
Figure~6 provides comparison cases between human-crafted item titles and our generated titles on the shopping platform.
Human-crafted titles are often informative but keyword-stuffed, lacking readability and user appeal.
In contrast, our generated titles are concise and stylistically diverse—ranging from benefit-driven to persona-targeted—using engaging language well-aligned with both item features and user attention, resulting in better CTR.

\section{Conclusion}
In this work, we propose CTOP, a novel CTR-driven ad text generation method via online feedback preference optimization.
CTOP employs a two-stage framework, comprising diverse ad text sampling strategy and CTR-driven preference optimization, tailored for the low-quality sampling-based decoding and noisy online feedback, respectively.
Extensive experiments have demonstrated the effectiveness of our method on both offline and online metrics, significantly improving CTR in real-world advertising scenarios.
This study contributes an effective solution to the industry and inspires further research in e-commerce advertising.

\bibliography{aaai2026}

\twocolumn[
\begin{center}
    {\LARGE \bfseries Supplementary Material}
\end{center}
\vspace{12ex}
]  

\appendix

\section{Appendices}
This supplementary material provides:
\begin{enumerate}
    \item \textbf{Exemplars from the High-Quality Ad Text Repository}: Illustrative cases showing several exemplars from the high-quality ad text repository in our diverse ad text sampling.
    \item \textbf{Details of Training Tasks}: Implementation details of the training tasks in our method, including the style transfer model, the reference model, and the policy model .
    \item \textbf{Additional Case Studies}: More detailed comparisons among human-crafted ad texts, underperforming candidates in diverse ad text sampling, and final outputs after our CTR-driven preference optimization, demonstrating the effectiveness of our proposed method.
\end{enumerate}

\section{Exemplars from the High-Quality Ad Text Repository}
In our diverse ad text sampling strategy, the high-quality ad text repository serves as a core component.
This repository consists of high-quality ad texts either crafted by domain experts or selected from successful online cases, with strong CTR performance proven beforehand.
Table~6 presents several exemplars from the repository.
Each exemplar in the repository details the item, the corresponding ad text, and the chain-of-thought (CoT) serving as imitation tips for style transfer.
The imitation tips provide a structural decomposition and a step-by-step writing guidance to capture the stylistic essence of the exemplar.
To address the inefficiency of online CoT during inference, we pre-generate offline CoTs for all exemplars in the repository.
Specifically, we use GPT-4 for CoT generation with the \texttt{Offline CoT Generation} prompt template (Table~8).
By introducing high-quality exemplars and their imitation tips as one-shot in-context examples, our strategy enables controllable style transfer and improves the diversity and effectiveness of candidate ad text generation.

\section{Details of Training Tasks}
Our method involves 3 models to be trained: (1) the \textbf{style transfer model} in diverse ad text sampling, (2) the \textbf{reference model} and (3) the \textbf{policy model} in CTR-driven preference optimization.
Table~6 presents the training details for each model used in our method, including the base model, the format, and the size of the training data.
Due to the business information involved in the training data, we are unable to publicly release the data used in our experiments.

The style transfer model is trained via supervised fine-tuning on instruction-response pairs.
The instruction is constructed by filling the \texttt{Style Transfer} prompt template (Table~8) with an exemplar ad text \texttt{\textless exemplar\textgreater} from the repository, its corresponding CoT  \texttt{\textless imitation tips\textgreater}, and the target item information \texttt{\textless item info\textgreater}.
The response is the generated ad text that imitates the style of the exemplar for the target item.
To construct the training data, we select 1,000 items disjoint from those in the high-quality ad text repository.
For each item, we randomly sample 10 exemplars and use GPT-4 to generate style-imitated ad texts using the same prompt template, resulting in 10K instruction-response pairs.
A Qwen-2.5-7B model is fine-tuned on this dataset for one epoch with a learning rate of $2 \times 10^{-5}$, yielding the final style transfer model.

The reference model is trained via supervised fine-tuning on instruction-response pairs to serve as a behavioral prior for subsequent preference optimization. 
The instruction is constructed by filling the \texttt{Ad Text Generation} prompt template (Table~8) with the target item information \texttt{\textless item info\textgreater}.
The response is the generated ad text for the target item.
We select 100K generated candidates in the diverse ad text sampling to construct the training data.
A Qwen-2.5-7B model is fine-tuned on this dataset for one epoch with a learning rate of $2 \times 10^{-5}$, yielding the reference  model.

The policy model, initialized from the reference model, is trained on preference pairs consisting of instruction $x$, preferred response $y_w$, and dispreferred response $y_l$.
The instruction is constructed by the same \texttt{Ad Text Generation} prompt template with the target item information \texttt{\textless item info\textgreater}.
We construct 500K preference pairs from the online A/B/n testing feedback, with each winning candidate of an item as $y_w$ and its human-crafted ad text as $y_l$.
Then, we leverage our CTR-driven preference optimization (Equation~7) to train the policy model for one epoch, with a learning rate of $2 \times 10^{-5}$ and scaling factor $m=2.5$ (Equation~5).

\section{Additional Case Studies}
Figure~7 presents detailed comparisons among human-crafted ad texts, underperforming candidates from diverse ad text sampling, and final outputs after our CTR-driven preference optimization. Human-crafted texts are informative but often lack persuasive elements such as emotional hooks. Diverse ad text sampling improves variation but may generate inappropriate candidates that fail to align with user preferences. In contrast, our method generates more attractive and targeted ad texts by learning high-CTR patterns from online feedback. For instance, it better utilizes bracketed emphasis (e.g., \textit{[IPX7 Waterproof]}) to highlight key selling points. Moreover, our model learns a better keyword-placing order to place attention-grabbing keywords or essential features earlier (e.g., \textit{[304 Stainless]} in the beginning). These cases demonstrate that our method generates ad texts that are well-aligned with both item features and user preferences, resulting in better CTR performance.

\begin{table*}[htbp]
\centering
\renewcommand{\arraystretch}{1.3}
\setlength{\tabcolsep}{6pt}
\begin{tabular}{@{}p{2.6cm}|p{6.6cm}|p{7.5cm}@{}}
\toprule
\multicolumn{1}{c|}{\textbf{Item}} & 
\multicolumn{1}{c|}{\textbf{Exemplar Ad Text}} & 
\multicolumn{1}{c}{\textbf{Imitation Tips (Chain-of-Thought)}} \\
\midrule
\texttt{Back Support Belt} &
\texttt{Say Goodbye to Back Pain Start with This Belt All-Day Comfort and Reliable Support} &
\ttfamily
\begin{minipage}{\linewidth}
\textbf{\#Structure} \\
targeted problem + action guidance + item name + additional information\\
\textbf{\#Writing Guidance}
\begin{itemize}[leftmargin=*,topsep=0pt,itemsep=0pt,parsep=0pt]
  \item Start with the specific problem or user pain point being addressed.
  \item Follow with an action-oriented guidance phrase.
  \item State the item name.
  \item Add supplementary information to highlight key features or advantages.
\end{itemize}
\end{minipage}\\
\midrule
\texttt{Shirt Dress} &
\texttt{Comfy\textbar{}Breathable\textbar{}Slimming-Vintage Print Shirt Dress} &
\ttfamily
\begin{minipage}{\linewidth}
\textbf{\#Structure} \\
benefit1\textbar{}2\textbar{}3-item description\\
\textbf{\#Writing Guidance}
\begin{itemize}[leftmargin=*,topsep=0pt,itemsep=0pt,parsep=0pt]
  \item Start with top 3 benefits using short keywords.
  \item List benefits in descending order of importance.
  \item Separate keywords with vertical bars "\textbar{}".
  \item Item description includes the item name and other key attributes.
  \item Use a dash "-" to connect the final benefit to the item description.
\end{itemize} 
\end{minipage}\\
\midrule
\texttt{Women's Heel} &
\texttt{[Petite Height Boost] Women's Chunky Heel, Waterproof Platform, French Design} &
\ttfamily
\begin{minipage}{\linewidth}
\textbf{\#Structure} \\
{[core benefit]} + item description\\
\textbf{\#Writing Guidance}
\begin{itemize}[leftmargin=*,topsep=0pt,itemsep=0pt,parsep=0pt]
  \item Start with the main selling point or targeted benefit, enclosed in brackets "[]".
  \item Item description includes the item name and other key attributes.
  \item Organize keywords of item description in descending order of importance.
\end{itemize}
\end{minipage}\\
\bottomrule
\end{tabular}
\caption{Illustrative exemplars from the high-quality ad text repository. Each exemplar lists the item, the ad text, and the imitation tips (CoT) for text style transfer, including a structural decomposition and a step-by-step writing guidance.}
\label{tab:adtextexamples}
\end{table*}

\begin{table*}[htbp]
\centering
\renewcommand{\arraystretch}{1.2}
\begin{tabular}{@{}c|c|c|c|c@{}}
\toprule
\textbf{Model} & \textbf{Base Model} & \textbf{Data Format} & \textbf{Template} & \textbf{Data Size} \\
\midrule
    Style Transfer Model & 
    Qwen-2.5-7B & 
    \texttt{instruction-response} & 
    \texttt{[Style Transfer]} & 
    10K \\
\midrule
    Reference Model & 
    Qwen-2.5-7B & 
    \texttt{instruction-response} & 
    \texttt{[Ad Text Generation]} & 
    100K \\
\midrule
    Policy Model & 
    Reference Model & 
    \texttt{preference pairs} $(x,y_w,y_l)$ & 
    \texttt{[Ad Text Generation]} & 
    500K \\
\bottomrule
\end{tabular}
\caption{Training details for each model used in our framework. The table summarizes for each model: the base model, data format used for training, the corresponding prompt template (see Table~8), and the data size.}
\label{tab:training_details}
\end{table*}

\begin{table*}[htbp]
\centering
\begin{tabular}{@{}p{2.3cm}|p{13.5cm}@{}}
\toprule
\multicolumn{1}{c|}{\textbf{Task}} & 
\multicolumn{1}{c}{\textbf{Prompt Template}} \\
\midrule
\makecell{\texttt{Offline CoT}\\\texttt{Generation}} &
\ttfamily
\begin{minipage}{\linewidth}
You are an experienced e-commerce ad specialist.

Given a exemplar ad text and its corresponding item information, please analyze the ad text and write some imitation tips for ad text style transfer.

\# Exemplar Ad Text\\
<exemplar>

\# Item Information\\
<item info>

\#Rules

- The imitation tips consist of a structural decomposition of the given ad text and a writing guidance for style transfer. \\
- Structural Decomposition: Provide an equation-like summary that reflects the overall structure of the ad text (e.g., benefit 1 + benefit 2 + item name + key attributes).\\
- Writing Guidance: Present, as an unordered list, actionable steps or suggestions on how to reproduce this style when generating new ad texts for other items.\\
- Output the result in the following format:

\# Structure\\
<structural decomposition>

\# Writing Guidance\\
<unordered guidance list>

\end{minipage}\\
\midrule
\makecell{\texttt{Style}\\\texttt{Transfer}} &
\ttfamily
\begin{minipage}{\linewidth}
You are an experienced e-commerce ad specialist. Please generate a new ad text for the target item by imitating the style of the given exemplar.

\# Exemplar ad text\\
<exemplar>

\# Style analysis\\
<imitation tips>

\# Target item information\\
<item info>

\# Notice\\
- Do not fabricate any details (e.g., price, promotion, shipping, endorsements) not present in item information.\\
- Output only the generated ad text without extra explanation.
\end{minipage}\\
\midrule
\makecell{\texttt{Ad Text}\\\texttt{Generation}}&
\ttfamily
\begin{minipage}{\linewidth}
You are an experienced e-commerce ad specialist capable of creating attractive ad texts that significantly enhance user interest and click-through rate. 
Please generate an ad text for the target item.

\# Target item information\\
<item info>

\# Notice\\
- Do not fabricate any details (e.g., price, promotion, shipping, endorsements) not present in item information.\\
- Output only the generated ad text without extra explanation.
\end{minipage}
\\
\bottomrule
\end{tabular}
\caption{Prompt templates for the offline CoT generation, the style transfer, and the ad text generation task.}
\label{tab:prompt-table}
\end{table*}

\begin{figure*}
    \centering
    \includegraphics[width=\linewidth]{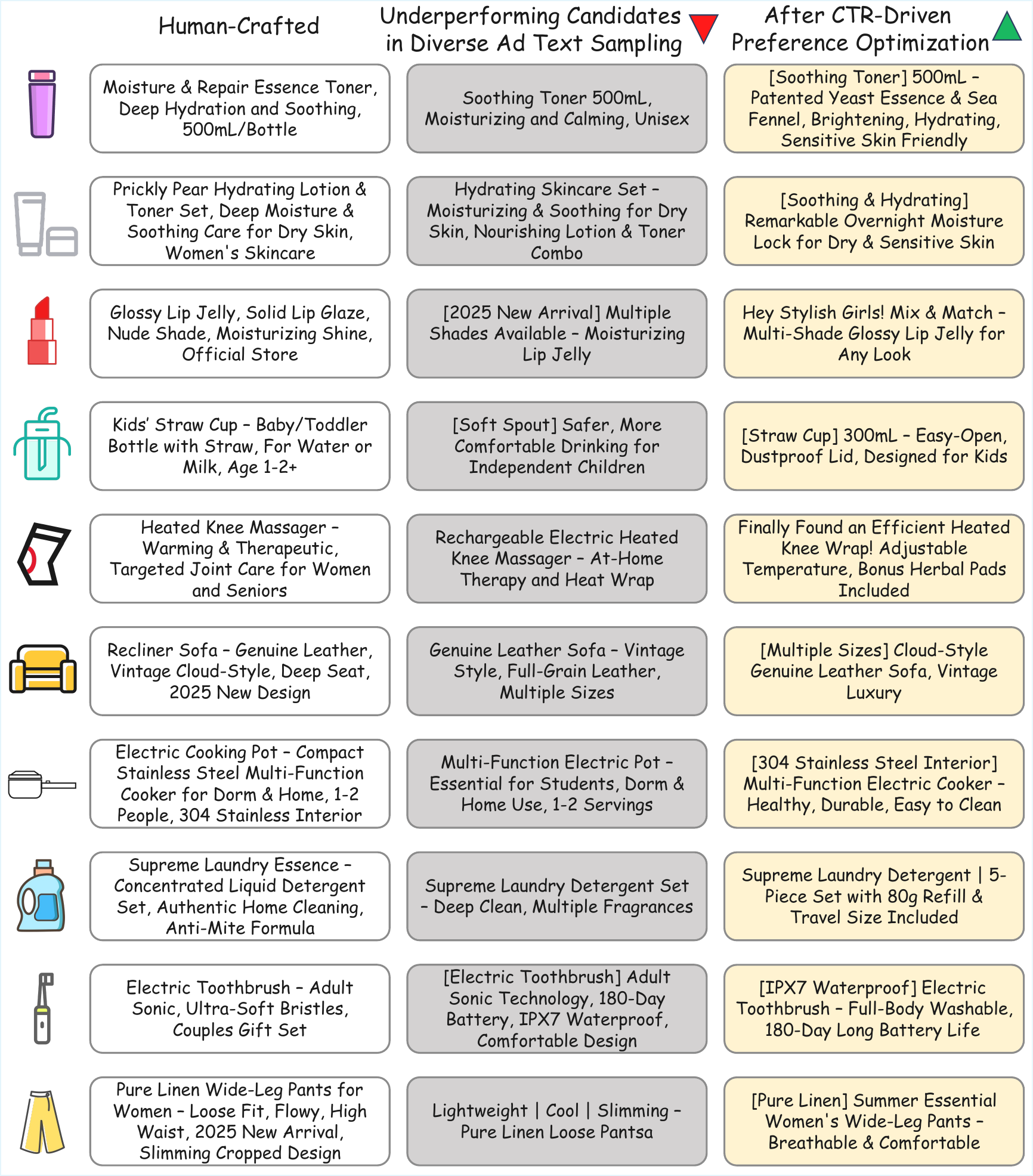}
    \caption{Detailed comparisons among human-crafted ad texts, underperforming candidates in diverse ad text sampling, and final outputs after our CTR-driven preference optimization.}
    \label{fig:more case study}
\end{figure*}

\end{document}